\documentclass[onecolumn,draftclsnofoot,peerreview,11pt]{IEEEtran}
\usepackage{}
\usepackage{bbm}

\usepackage{amsmath}
\usepackage{amssymb}
\usepackage{tabularx}

\usepackage{graphicx}

\usepackage{hyperref}

\usepackage{threeparttable}
\usepackage{dcolumn}
\usepackage{multirow}
\usepackage{booktabs}
\usepackage{algpseudocode}
\usepackage{algorithm}
\usepackage{indentfirst}

\ifCLASSINFOpdf
\else
\fi

\begin{document}
%
\title{Layered Subspace Codes for Network Coding}

\author{Chao Chen, Hongmei Xie, and Baoming Bai \thanks{Chao Chen is with China Academy of Space Technology
(Xi'an), Xi'an, China, 710000.  E-mail: chenchaoxidian@gmail.com }
\thanks{Hongmei Xie is with the Department of Electrical and Computer
Engineering, Lehigh University, Bethlehem, PA 18015, USA. E-mail:
hox209@lehigh.edu}
\thanks{Baoming Bai is with State Key Lab of ISN, Xidian University, Xi'an, China,
710071.  E-mail: bmbai@mail.xidian.edu.cn} }



\maketitle


\begin{abstract}
Subspace codes were introduced by K\"otter and Kschischang for error
control in random linear network coding. In this paper, a layered
type of subspace codes is considered, which can be viewed as a
superposition of multiple component subspace codes. Exploiting the
layered structure, we develop two decoding algorithms for these
codes. The first algorithm operates by separately decoding each
component code. The second algorithm is similar to the successive
interference cancellation (SIC) algorithm for conventional
superposition coding, and further permits an iterative version. We
show that both algorithms decode not only deterministically up to
but also probabilistically beyond the error-correction capability of
the overall code. Finally we present possible applications of
layered subspace codes in several network coding scenarios.


\end{abstract}



\section{Introduction}

In the paradigm of network coding \cite{R1}, information
transmission is highly susceptible to packet errors. Due to the
mixing nature, even a single corrupt packet may cause widespread
error propagation, rendering the entire transmission useless. Thus
error control is essential for providing reliable transmission in
network coding.

The notion of network error correction coding was first introduced
by Cai and Yeung in \cite{R3}. Their approach is based on a coherent
transmission model, in which both the transmitter and receiver know
the network topology.

In the context of random linear network coding \cite{R2}, K\"otter
and Kschischang proposed the subspace coding method as the error
control solution \cite{R4}. A noncoherent transmission model was
assumed where neither the transmitter nor receiver have knowledge of
the network topology and the particular network codes used. Subspace
codes encapsulate network codes to provide an end-to-end error
protection.

Recently, a coding scheme consisting of a number of subspace codes
was proposed by Siavoshani et al. for multi-source multicast network
coding \cite{R7}. In \cite{R8}, Dikaliotis et al. extended this work
by constructing capacity-approaching subspace coding schemes for
multi-source network coding transmission.

In this paper, we investigate the superposition property of the
codes in \cite{R7} and propose two decoding algorithms. Due to their
layered structure, we refer to the codes as layered subspace codes.
Our main contributions can be summarized as follows.
\begin{itemize}
\item We provide more insights by showing that a layered subspace code forms a superposition coding
scheme \cite{R10}.
\item We develop two efficient decoding algorithms. The first
algorithm operates by separately decoding each component code. The
second algorithm is similar to the successive interference
cancellation (SIC) algorithm for conventional superposition coding,
and further permits an iterative version. We show that both
algorithms are guaranteed to decode up to the error-correction
capability of the overall code. Besides, they can occasionally
decode beyond the capability.
\item We point out that layered subspace codes can find more applications than
presented in \cite{R7}. For example, the codes can be used as an
adaptive transmission scheme or an unequal error protection scheme
for single-source multicast network coding.
\end{itemize}

The rest of the paper is organized as follows. Section II gives a
brief review of subspace codes. In Section III, we investigate the
properties of layered subspace codes and develop two decoding
algorithms for these codes. Section IV discusses some possible
applications of layered subspace codes in network coding. Finally,
we conclude the paper in Section V.

\section{Preliminaries}
In this section, we briefly recall the subspace coding method
\cite{R4} for random linear network coding (RLNC) \cite{R2}. In
RLNC, a source injects some packets into the network, each being
regarded as a row vector over a given finite field. These packets
propagate though the network, passing though a number of
intermediate nodes between source and receiver. Each intermediate
node creates a random linear combination of packets it received, and
transmits this combination. Finally, a receiver collects a set of
such randomly generated packets and tries to recover the packets
injected into the network.

Let $F_{q}$ be a finite field with $q$ elements, where $q$ is a
prime power. Let $W$ be a fixed finite-dimensional vector space over
$F_{q}$ and $\mathcal {P}(W)$ the set of all subspaces of $W$.
Denote by dim$(V)$ the dimension of an element $V \in \mathcal
{P}(W)$. Two operations on $\mathcal {P}(W)$ can be defined
\cite{R11}. The intersection of $V, U \in \mathcal {P}(W)$ is
defined as
\begin{align}
V \cap U \doteq \{w: w \in V, w \in U\},
\end{align}
which is the subspace of largest dimension contained in both $V$ and
$U$. The sum of $V$ and $U$ is defined as
\begin{align}
V + U \doteq \{v+u: v \in V, u \in U\},
\end{align}
which is the subspace of smallest dimension containing both $V$ and
$U$. If $V$ and $U$ intersect trivially (i.e., $V \cap U =\{0\}$),
$V + U$ is called the direct sum, denoted by $V\oplus U$.

For RLNC, the transmission is modeled as an operator channel, where
both the input and output are a subspace of $W$ \cite{R4}. Let $V$
be the input and $U$ the output, the operator channel relates them
by
\begin{align}
U = (V \cap U) \oplus E,
\end{align}
where $E$ is called the error space. In transforming from $V$ to
$U$, it is said that the operator channel commits $\rho =
\textrm{dim}(V)-\textrm{dim}(V \cap U)$ erasures (also called
deletions of dimension) and $t = \textrm{dim}(E) =
\textrm{dim}(U)-\textrm{dim}(V \cap U)$ errors (also called
insertions of dimension). In practice, the source sends a basis for
the information-carrying vector space $V$ and the receiver collects
a set of vectors that span the possibly corrupt vector space $U$.

To measure the degree of dissimilarity between $V$ and $U$, the
subspace distance has been introduced \cite{R4}
\begin{align}
d_{S}(V,U) & \doteq \textrm{dim}(V+U) - \textrm{dim}(V \cap U) \nonumber \\
 &=\textrm{dim}(V) +\textrm{dim}(U) -2\textrm{dim}(V \cap U)
\nonumber \\
&= \rho + t.
\end{align}
With the definition, $\mathcal{P}(W)$ forms a metric space.

A subspace code $\mathcal {C}$ is defined to be a nonempty subset of
$\mathcal {P}(W)$ \cite{R4}. Each codeword of $\mathcal {C}$ is a
subspace of $W$. The minimum (subspace) distance of $\mathcal {C}$
is defined as
\begin{align}
d_{S}(\mathcal {C}) \doteq \min_{V,V'\in \mathcal {C}:V \neq
V'}{d_{S}(V,V')}.
\end{align} A subspace code with minimum distance $d_{S}(\mathcal
{C})>2(\rho+t)$ is capable of correcting any $\rho$ erasures and $t$
errors with the minimum distance decoder. That is, if $2d_{S}(V,U) <
d_{S}(\mathcal {C})$, the transmitted $V$ can be recovered from the
received $U$.

One major construction of subspace codes \cite{R5} is through
lifting the so called rank-metric codes \cite{R6}. Let $F_{q}^{n
\times m}$ be the set of all $n \times m$ matrices over $F_{q}$. For
$X, Y \in F_{q}^{n \times m}$, the rank distance between $X$ and $Y$
is defined as
\begin{align}
d_{R}(X,Y) \doteq \textrm{rank}(X-Y).
\end{align}
 A rank-matric code $\mathcal {M}$ is defined to be a nonempty
 subset of $F_{q}^{n \times m}$. Each codeword of $\mathcal {M}$ is
 a $n \times m$ matrix over $F_{q}$. The minimum (rank) distance of $\mathcal {M}$
 is defined as
\begin{align}
d_{R}(\mathcal {M}) \doteq \min_{X,X'\in \mathcal {M}:X \neq
X'}{d_{R}(X,X')}.
\end{align}
The most well-known rank-metric codes are Gabidulin codes \cite{R6},
which have the maximum possible minimum rank-distance, analogous to
the Reed-Solomon codes in Hamming metric.

Let $I_{n}$ be the $n \times n$ identity matrix. Denote by $\langle
X \rangle$ the vector space spanned by rows of a matrix $X$ over
$F_{q}$. The lifting of a rank-metric code $\mathcal {M}$ gives the
subspace code
\begin{align}
\setlength{\arraycolsep}{2.3pt} \mathcal {C} \doteq \left\{V:
V=\left \langle \left[
\begin{array}{cc}
I_{n} & X
\end{array}
\right] \right \rangle , X \in \mathcal {M}\right\}.
\end{align}
It can be proved that $d_{S}(\mathcal {C})=d_{R}(\mathcal {M})$
\cite{R5}. If $\mathcal {M}$ is a Gabidulin code, two efficient
decoding algorithms have been developed for the resulting subspace
code \cite{R4}, \cite{R5}. Both algorithms are guaranteed to decode
up to the error-correction capability of the subspace code.

\section{Layered Subspace Codes}

\subsection{Code description}

Let $\mathcal {M}_{l} \subseteq F_{q}^{n_{l} \times m} \,
(l=1,2,\cdots,L)$ be $L$ rank-metric codes. We define the overall
subspace code as
\begin{align}
\renewcommand\arraystretch{1.1}
\setlength{\arraycolsep}{2.5pt} \mathcal {C} \doteq  \left\{V: V=
\left< \left[\begin{array}{ccccc} I_{n_{1}} & 0  & \cdots
&0 & X_{1}\\
 0 &  I_{n_{2}} & \cdots
&0 & X_{2}\\
\vdots & \vdots  & \ddots
&\vdots & \vdots \\
0 & 0  & \cdots &I_{n_{L}} & X_{L}
\end{array} \right] \right>, X_{1}\in \mathcal {M}_{1}, \cdots, X_{L}\in \mathcal {M}_{L}
\right\}.
\end{align}
By lifting the rank-metric codes, we can obtain $L$ component
subspace codes
\begin{align}
\setlength{\arraycolsep}{3pt} \mathcal {C}_{l} \doteq  \left\{
V_{l}: V_{l}=  \left< \left[\begin{array}{cccc} 0_{l_{1}} &
I_{n_{l}} & 0_{l_{2}} & X_{l}
\end{array} \right] \right>,  X_{l} \in \mathcal {M}_{l}
\right\} \,\,\,\, (l=1,2\cdots,L),
\end{align}
where $0_{l_{1}}$ is the $n_{l} \times \left(\sum_{i=1}^{l-1}{n_{i}}
\right)$ all-zero matrix and $0_{l_{2}}$ is the $n_{l} \times
\left(\sum_{i=l+1}^{L-1}{n_{i}}\right)$ all-zero matrix. For
decoding purpose, we will assume the rank-metric codes to be
Gabidulin codes.


Obviously, for any $V_{i} \in \mathcal {C}_{i}$ and $V_{j} \in
\mathcal {C}_{j}$ $(i \neq j)$, $V_{i} \cap  V_{j} =\{0\}$.
Therefore, we have

\emph{Property 1:} $\mathcal {C}=\left\{V: V=V_{1}\oplus V_{2}\oplus
\cdots \oplus V_{L}, V_{l} \in \mathcal {C}_{l}\right\}. $

The property leads to a superposition coding scheme, which is
depicted in Fig. 1. The overall subspace code consists of $L$
superimposed layers (each corresponding to a component code), and
hence we have the name layered subspace code.

\begin{figure}[!ht]
 \centering
\includegraphics[width=3.2in]{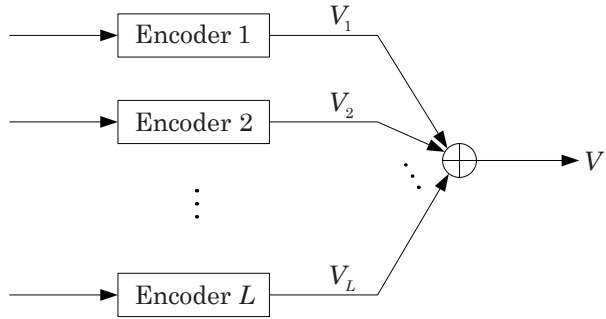}
\caption{The overall subspace code as a superposition coding scheme.} 
\end{figure}

Based on the definition (9) and (10), we further have the following
property.

\emph{Property 2:} For any two codewords of $\mathcal {C}$,
$V=V_{1}\oplus  \cdots \oplus V_{L} \,\, (V_{l} \in \mathcal
{C}_{l})$ and $V^{'}=V_{1}^{'}\oplus  \cdots \oplus V_{L}^{'} \,\,
(V_{l}^{'} \in \mathcal {C}_{l}^{'})$, $V=V^{'}$ if and only if
$V_{l} = V_{l}^{'}$ for all $l$.

For the minimum distances of $\mathcal {C}$ and $\mathcal {C}_{l}$,
the following property holds. For the proof, see \cite{R6}.

\emph{Property 3:} $d_{S}(\mathcal {C}) = \min\big\{d_{S}(\mathcal
{C}_{1}),d_{S}(\mathcal {C}_{2}),\cdots,d_{S}(\mathcal
{C}_{L})\big\}$.

\subsection{Decoding algorithm I}

Suppose that a codeword $V \in \mathcal {C}$ was transmitted and the
vector space $U$ is now received. Corresponding to each $V_{l}
\subseteq V$, we define a subspace $U_{l} \subseteq U$ as follows.
It consists of all vectors of $U$ such that the elements at the
coordinates $\big\{1,\cdots,\sum_{i=1}^{l-1}{n_{i}} \big\}\cup
\big\{\sum_{i=1}^{l}{n_{i}+1, \cdots, \sum_{i=1}^{L}n_{i} }\big\}$
are zero. Now, we can describe decoding algorithm I as follows:
\begin{enumerate}
\item[1)] Extract $U_{l}$ from $U$;
\item[2)] Use the decoder for $\mathcal {C}_{l}$ to recover $V_{l}$
from $U_{l}$ \cite{R4}, \cite{R5}.
\end{enumerate}

Note that given a set of vectors that span $U$, a basis of $U_{l}$
can be extracted with the aid of Gauss-Jordan elimination. Based on
Property 2, once $V_{l} \in \mathcal {C}_{l}$ for all $l$ can be
recovered, the transmitted codeword $V \in \mathcal {C}$ can be
determined as $V=V_{1}\oplus \cdots \oplus V_{L}$.

\begin{figure}[!ht]
 \centering
\includegraphics[width=3.5in]{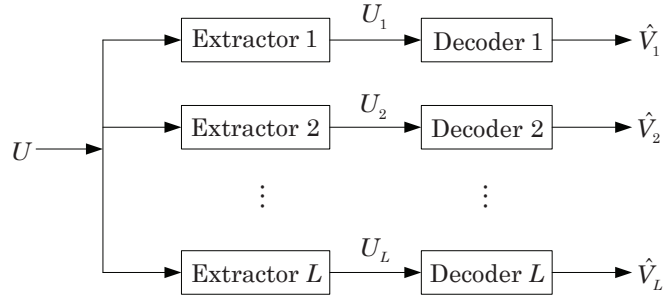}
\caption{The schematic diagram of decoding algorithm I.} 
\end{figure}

An illustration of the decoding algorithm is given in Fig. 2. It is
seen that a parallel implementation is allowed. Moreover, if a
receiver is only interested in a particular $V_{l}$, then he only
needs to perform the corresponding layer in Fig. 2.

We now focus on the error-correction ability of decoding algorithm
I. We first need to introduce a general result.

\emph{Lemma 1:} Let $A$ and $B$ be two subspaces of $W$. Let $A^{'}$
be a subspace of $A$. Then,
\begin{align}
\textrm{dim}(A) - \textrm{dim}(A\cap B) \geqslant
\textrm{dim}(A^{'}) -\textrm{dim}(A^{'} \cap B).
\end{align}

\hspace{0.3cm}\emph{Proof:} Since $A\cap B$ is a subspace of $A$,
there exists a (not unique\footnote{For an arbitrary $w \neq 0 \in
A\cap B$, we have $C^{'}\doteq \{c+w:c\in C\} \neq C$ such that $A =
(A \cap B) \oplus C^{'}$. Note that all such vector spaces are
isomorphic to the quotient space $A\backslash (A\cap B)$
\cite{R11}.}) subspace $C$ of $A$ such that $A = (A \cap B) \oplus
C$. Similarly, there exists a (not unique) subspace $D$ of $A^{'}$
such that $A^{'} = (A^{'} \cap B) \oplus D$. Therefore, we only need
to prove $\textrm{dim}(C) \geqslant \textrm{dim}(D)$.

Assume that there exists an $x \neq 0 \in D\cap (A \cap B)$. Then $x
\in D$ and $x \in B$. Since $A^{'} = (A^{'} \cap B) \oplus D$, we
have $x \in A^{'}$. So $x \in A^{'}\cap B$, which together with $x
\in D$ contradicts the fact $(A^{'}\cap B)\cap D=\{0\}$. Therefore,
$D \cap (A \cap B) = \{0\}$. Since $A^{'}$ is a subspace of $A$ (by
hypothesis) and $A^{'} = (A^{'} \cap B) \oplus D$, $D$ is a subspace
of $A$. So $(A\cap B)\oplus D$ is a subspace of $A=(A\cap B)\oplus
C$. Therefore, $\textrm{dim}(C) \geqslant \textrm{dim}(D)$,  and the
statement of the lemma follows. \hspace{11cm} $\square$

For decoding algorithm I, we have the following result.

\emph{Theorem 2:} $d_{S}(U,V) \geqslant d_{S}(U_{l},V_{l})$ for all
$l$.

\hspace{0.3cm}\emph{Proof:} It is important to note that
\begin{align}
V_{l} \cap U = V_{l} \cap U_{l}
\end{align}
and
\begin{align}
U_{l} \cap V = U_{l} \cap V_{l}.
\end{align}
Then, based on Lemma 1, we have
\begin{align}
\textrm{dim}(V) - \textrm{dim}(V\cap U) \geqslant
\textrm{dim}(V_{l}) - \textrm{dim}(V_{l}\cap U_{l})
\end{align}
and
\begin{align}
\textrm{dim}(U) - \textrm{dim}(U\cap V) \geqslant
\textrm{dim}(U_{l}) - \textrm{dim}(U_{l}\cap V_{l}).
\end{align}
Summing up (14) and (15), we obtain
\begin{align}
\textrm{dim}(V) + \textrm{dim}(U)- 2\textrm{dim}(U\cap V) \geqslant
  \textrm{dim}(V_{l}) + \textrm{dim}(U_{l}) -
2\textrm{dim}(U_{l}\cap V_{l}).
\end{align}
By the definition of subspace distance, $d_{S}(V,U) \geqslant
d_{S}(V_{l},U_{l})$. \hspace{6cm} $\square$

Combining Property 3 and Theorem 2, we have

\emph{Corollary 3:} If $2d_{S}(V,U) < d_{S}(\mathcal {C})$, then
$2d_{S}(V_{l},U_{l}) < d_{S}(\mathcal {C}_{l})$ for all $l$.

The corollary indicates that decoding algorithm I is guaranteed to
decode up to the error-correction capability of the overall code.

Note that it may happen that $2d_{S}(V,U) \geqslant d_{S}(\mathcal
{C})$ while $2d_{S}(V_{l},U_{l})<d_{S}(\mathcal {C}_{l})$. In this
case, the decoding algorithm can decode beyond the error-correction
capability of the overall code. We give an example to show this.

\begin{figure}[!ht]
 \centering
\includegraphics[width=5.8in]{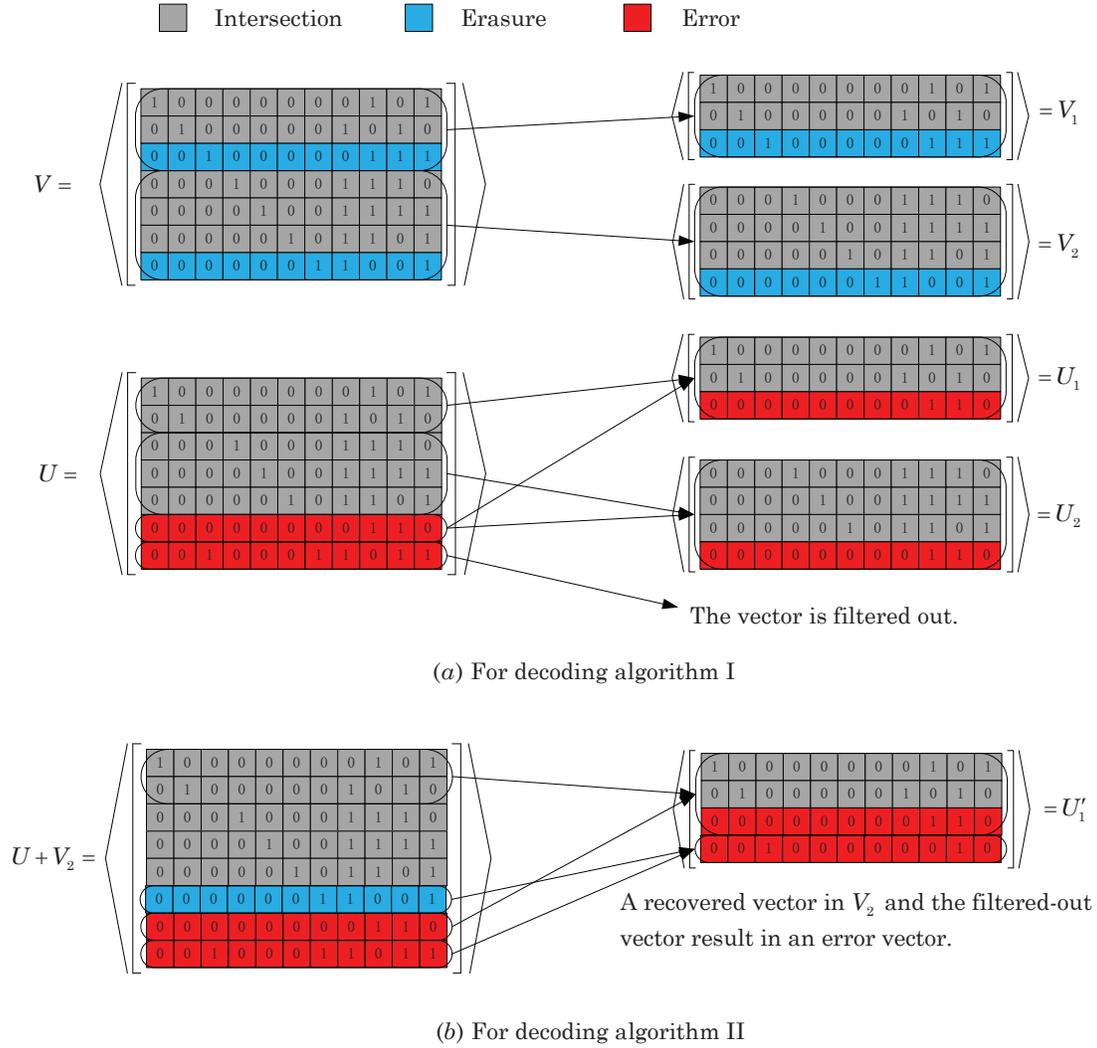}
\caption{A transmitted codeword $V \in \mathcal {C}$ and the corresponding received $U$.} 
\end{figure}

\emph{Example 1:} Let the overall code $\mathcal {C}$ be composed of
two component codes $\mathcal {C}_{1}$ and $\mathcal {C}_{2}$, which
have parameters $n_{1} = 3$, $n_{2}=4$, $m=4$, $d_{S}(\mathcal
{C}_{1})=6$ and $d_{S}(\mathcal {C}_{2})=8$. According to Property
3, we have $d_{S}(\mathcal {C})=6$. Fig. 3 (a) gives a transmitted
codeword $V \in \mathcal {C}$ and the corresponding received vector
space $U$. Also shown in the figure are $V_{1} \in \mathcal
{C}_{1}$, $V_{2} \in \mathcal {C}_{2}$, and $U_{1}$ and $U_{2}$
extracted from $U$. We note that in extracting $U_{1}$ and $U_{2}$,
the vector of $U$ in the last row is filtered out.

It is easily verified that $d_{S}(V,U)=7+7-2\times 5=4$,
$d_{s}(V_{1},U_{1})=3+3-2 \times 2 = 2$, and
$d_{S}(V_{2},U_{2})=4+4-2 \times 3=2$. Consequently, we have
$2d_{S}(V,U)>d_{S}(\mathcal {C})$,
$2d_{S}(V_{1},U_{1})<d_{S}(\mathcal {C}_{1})$, and
$2d_{S}(V_{2},U_{2})<d_{S}(\mathcal {C}_{2})$. Therefore, for the
instantiated $V$ and $U$, decoding algorithm I decodes beyond the
error-correction capability of the overall code.  \hspace{11.5cm}
$\square$

\subsection{Decoding algorithm II}

It is well-known that for conventional superposition coding, the
(iterative) successive interference cancellation (SIC) decoding
algorithm is usually adopted \cite{R10}. Viewing a layered subspace
code as a superposition coding scheme, we develop a SIC-like
decoding algorithm, which is shown in Fig. 4. With a slight abuse of
notation, the `$\oplus $' here denotes the sum of two vector spaces
that do not necessarily intersect trivially. By taking the dashed
arrows into account, we obtain an iterative version of the
algorithm. When decoder $l$ does not decode into a codeword of
$\mathcal {C}_{l}$ (this can be checked by the decoder), we set
$\hat{V}_{l}$ to be the zero subspace. So if there only occur
erasures in the operator channel, the iterative version in general
outperforms its non-iterative counterpart.

\begin{figure}[!ht]
 \centering
\includegraphics[width=4.1in]{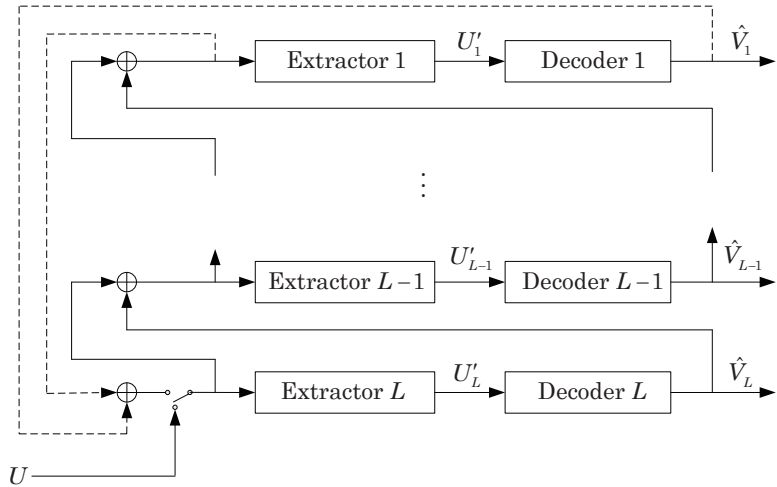}
\caption{The schematic diagram of decoding algorithm II (dashed arrows for the iterative version).} 
\end{figure}

It should be pointed out that in \cite{R8}, the authors have used
the idea of SIC to decode their constructed subspace codes. However,
they did not mention any iterative decoding.

On the error-correction ability of decoding algorithm II, we have
the following result.

\emph{Theorem 4:} If $2d_{S}(V,U) < d_{S}(\mathcal {C})$, then
$d_{S}(V,U)\geqslant d_{S}(V,U+\hat{V}_{L}) \geqslant \cdots
\geqslant d_{S}(V,U+\hat{V}_{L}+\cdots+\hat{V}_{1})$.

\hspace{0.3cm}\emph{Proof:} We prove the theorem by induction on
$l$. Since $2d_{S}(V,U) < d_{S}(\mathcal {C})$, from Corollary 3, we
have $\hat{V}_{l}=V_{l} \subseteq V$. Therefore,
$\textrm{dim}(V+U)=\textrm{dim}\big(V+(U+\hat{V}_{l})\big)$ and
$\textrm{dim}(V\cap U)\leqslant \textrm{dim}\big(V\cap
(U+\hat{V}_{l})\big)$. Based on the definition of subspace distance,
we have $d_{S}(V,U)\geqslant d_{S}(V,U+\hat{V}_{L})$.

Assume that $d_{S}(V,U)\geqslant d_{S}(V,U+\hat{V}_{L}) \geqslant
\cdots \geqslant d_{S}(V,U+\hat{V}_{L}+\cdots+\hat{V}_{l})$ holds.
Since $2d_{S}(V,U) < d_{S}(\mathcal {C})$,
$2d_{S}(V,U+\hat{V}_{L}+\cdots+\hat{V}_{l}) < d_{S}(\mathcal {C})$.
From Fig. 4, we see that $U+\hat{V}_{L}+\cdots+\hat{V}_{l}$ is the
input to the extractor $l-1$. Based on Corollary 3, we have
$\hat{V}_{l-1}=V_{l-1} \subseteq V$. Consequently,
$d_{S}(V,U+\hat{V}_{L}+\cdots+\hat{V}_{l}) \geqslant
d_{S}(V,U+\hat{V}_{L}+\cdots+\hat{V}_{l-1})$. Thereby, the proof is
complete. \hspace{6.5cm} $\square$

From the proving process, we see that decoding algorithm II is
guaranteed to decode up to the error-correction capability of the
overall code.

Like decoding algorithm I, decoding algorithm II also occasionally
decodes beyond the error-correction capability of the overall code.
We note that the two algorithms may correct different errors in this
case. We show this through the following example.

\begin{figure}[!ht]
 \centering
\includegraphics[width=5.9in]{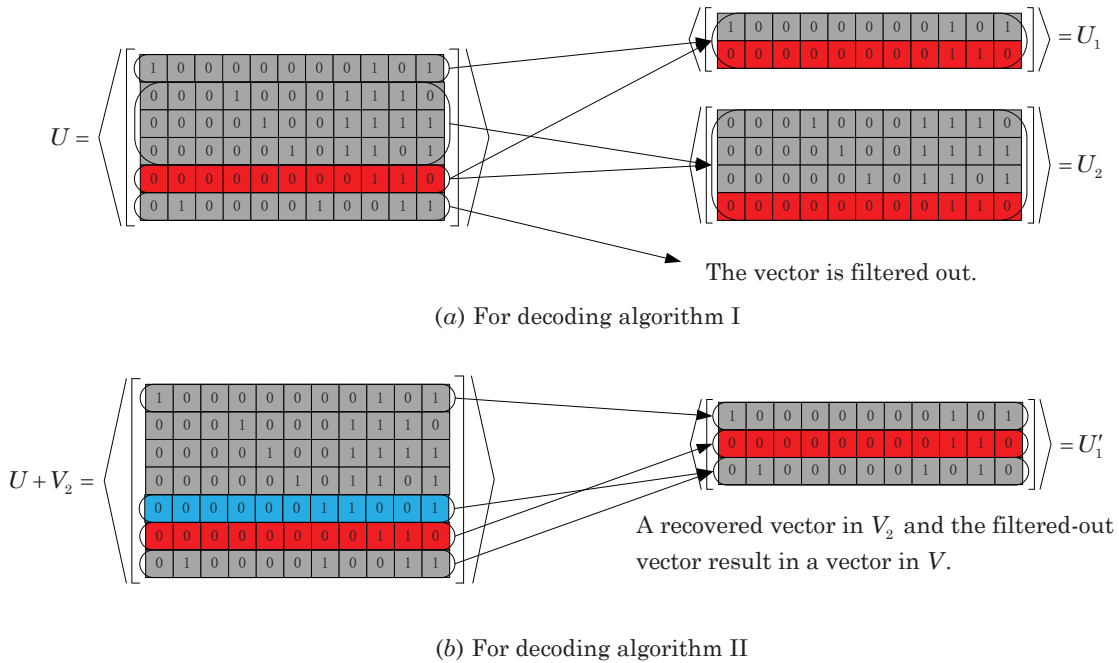}
\caption{Decoding for another received $U$.} 
\end{figure}

\emph{Example 2:} For the parameters given in Example 1, we now use
decoding algorithm II. Since $U_{2}^{'}=U_{2}$, $V_{2}$ can be
recovered. The resulting $U+V_{2}$ and $U_{1}^{'}$ is shown in Fig.
3 (b). It is easily obtained that $d_{S}(V_{1},U_{1}^{'})=3+4-2
\times 2=3$. Since $2d_{S}(V_{1},U_{1}^{'})=d_{S}(\mathcal
{C}_{1})$, $V_{1}$ cannot be recovered with decoding algorithm II.

Consider again the transmitted $V$ in Example 1, but now suppose
that the received $U$ is given as in Fig. 5. It is easily verified
that $d_{S}(V,U)=7+6-2\times 5=3$, $d_{S}(V_{1},U_{1})=3+2-2 \times
1 = 3$, and $d_{S}(V_{2},U_{2})=4+4-2 \times 3=2$. Therefore, with
decoding algorithm I, only $V_{2}$ can be recovered.

From the recovered $V_{2}$, we calculate $U+V_{2}$ and $U_{1}^{'}$
as in Fig. 5 (b). Since $d_{S}(V_{1},U_{1}^{'})=3+3-2 \times 2 = 2$,
we have $2d_{S}(V_{1},U_{1}^{'}) < d_{S}(\mathcal {C})$. So $V_{1}$
can be recovered with decoding algorithm II.

In summary, for the transmitted $V$ and the received $U$ in Fig. 3,
only decoding algorithm I can correctly decode, while for the same
$V$ and another $U$ as given in Fig. 5, only decoding algorithm II
can correctly decode.  \hspace{14.9cm}$\square$

\section{Applications}
In this section, we show that layered subspace codes can be applied
in various scenarios for network coding.

\begin{figure}[!ht]
 \centering
\includegraphics[width=6.3in]{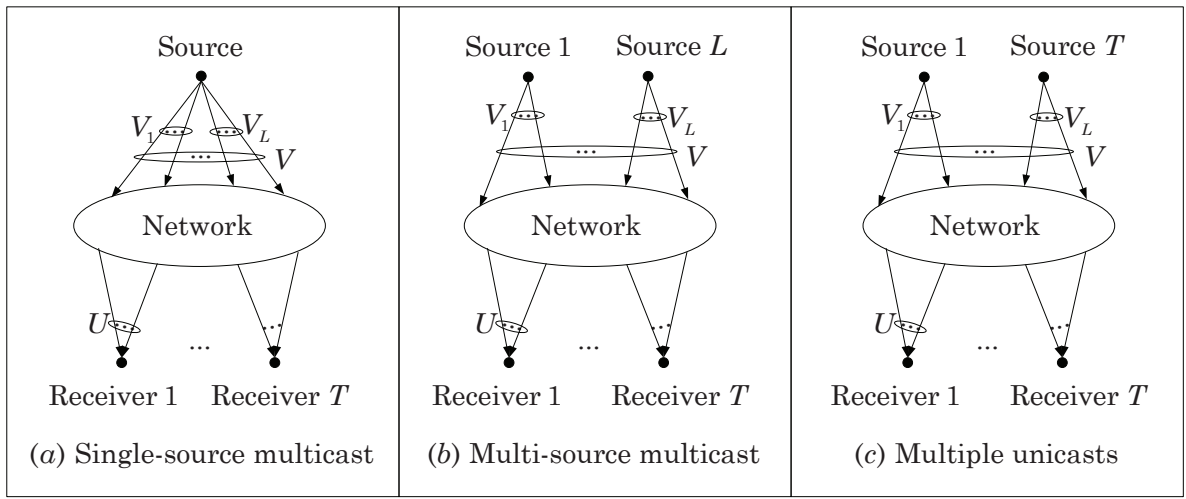}
\caption{Possible application scenarios of layered subspace codes.} 
\end{figure}

\begin{itemize}
\item Single-source multicast

   \setlength{\parindent}{1em}In this scenario, a single source communicates its information
   over a network to a specified set of $T$ receivers, as shown in Fig.
   6 (a). The source is encoded with the overall subspace code and the
   basis vectors defining a codeword $V \in \mathcal {C}$ in (9) are
   transmitted. To deal with network dynamics, the number of
   component codes can be adapted. Therefore, this leads to an
   adaptive transmission scheme. On the other hand, by using
   component codes with different error-correction capabilities, the
   coding scheme can be used for unequal protection transmission \cite{R9}.
\item Multi-source multicast \cite{R7}

  \setlength{\parindent}{1em} As shown in Fig. 6 (b), $L$ sources
  transmit independent information over a network to a
  specified set of $T$ receivers. Source $l$ is encoded with
  component code $l$ and the basis vectors defining a codeword $V_{l} \in \mathcal
  {C}_{l}$ in (10) are transmitted. Based on the received vectors,
  each receiver tries to recover $V_{l}$ for all $l$.
\item Multiple unicasts

  \setlength{\parindent}{1em} A unicast means that a single source
  communicates its information over a network to a single receiver.
  In the multiple unicasts scenario, the number of receivers is
  equal to that of sources, as shown in Fig. 6 (c), and receiver $l$
  only requests the information from source $l$. The coding scheme
  is the same as in the multi-source multicast scenario. Since
  receiver $l$ only wishes to recover $V_{l}$, decoding algorithm I
  is preferred in this scenario.
\end{itemize}

\section{Conclusion}
We treated the layered subspace codes in \cite{R7} as a
superposition coding scheme and proposed two efficient decoding
algorithms. Error-correction abilities of both algorithms are
analyzed. As an error control scheme, layered subspace codes can be
expected to find various applications for network coding.



\section*{Acknowledgment}
This work was jointly supported by NSFC grant 61101127 and the 973
Program of China 2012CB316100.

\end{document}